\documentclass{article}

\usepackage[twocolumn,margin=17mm,columnsep=12mm]{geometry}
\usepackage{multicol}

\usepackage{authblk}

\usepackage[compress]{natbib}
\setcitestyle{super}
\bibpunct{}{}{,}{s}{}{\textsuperscript{,}}

\usepackage{abstract}

\usepackage{lineno}

\usepackage{sectsty}
\allsectionsfont{\large\textsf}

\usepackage[dvipsnames]{xcolor}

\usepackage{float}
\usepackage{amsmath}
\usepackage{amssymb}
\usepackage[utf8x]{inputenc}
\usepackage[T1]{fontenc}
\usepackage[english]{babel}
\usepackage{mathptmx}
\usepackage{fancyhdr}
\usepackage{braket}
\usepackage{graphicx}
\usepackage{enumerate}

\usepackage{epstopdf}
\usepackage[squaren]{SIunits}

\usepackage[colorlinks=true, linkcolor=blue, urlcolor=blue, citecolor=blue, plainpages=false, pdfpagelabels, bookmarksnumbered]{hyperref}
\usepackage{todonotes}

\def\XXint#1#2#3{{\setbox0=\hbox{$#1{#2#3}{\int}$}
    \vcenter{\hbox{$#2#3$}}\kern-.5\wd0}}

\renewcommand{\vec}[1]{\mathbf{#1}}

\newcommand{\bmat}[1]{
  \begin{bmatrix}
    #1
  \end{bmatrix}
}

\newenvironment{methods}{%
  \setlength{\parindent}{0in}%
  \small%
  \section*{Methods}%
  \setlength{\parskip}{12pt}%
}{}

\newenvironment{addendum}{%
  \setlength{\parindent}{0in}%
  \small%
  \begin{list}{Acknowledgements}{%
      \setlength{\leftmargin}{0in}%
      \setlength{\listparindent}{0in}%
      \setlength{\labelsep}{0em}%
      \setlength{\labelwidth}{0in}%
      \setlength{\itemsep}{12pt}%
      }
  }
  {\end{list}\normalsize}

\begin{document}
\makeatletter

\title{Sub-cycle ionization dynamics revealed by trajectory resolved,
  elliptically-driven high-order harmonic generation}
\date{}

\author[1]{E.~W.~Larsen\thanks{esben.witting-larsen@fysik.lth.se}}
\affil[1]{Department of Physics, Lund University, P.O. Box 118, SE-221 00 Lund, Sweden.}
\author[1]{S.~Carlstr\"om}
\author[1]{E.~Lorek}
\author[1]{C.~M.~Heyl}
\author[2,3]{D.~Pale\v{c}ek}
\affil[2]{Department of Chemical Physics, Lund University, P.O. Box 124, SE-22100 Lund, Sweden.}
\affil[3]{Department of Chemical Physics, Charles University in Prague, Ke Karlovu 3, 121 16 Praha 2, Czech Republic.}
\author[4]{K.~J.~Schafer}
\affil[4]{Louisiana State University, Baton Rouge, 70803-4001, Louisiana, United States of America.}
\author[1]{A.~L'Huillier}
\author[2]{D.~Zigmantas}
\author[1]{J.~Mauritsson.}

\makeatother

\twocolumn[{
\maketitle
\begin{onecolabstract}

  The sub-cycle dynamics of electrons driven by strong laser fields is
  central to the emerging field of attosecond science. We demonstrate
  how the dynamics can be probed through high-order harmonic
  generation, where different trajectories leading to the same
  harmonic order are initiated at different times, thereby probing
  different field strengths. We find large differences between the
  trajectories with respect to both their sensitivity to driving field
  ellipticity and resonant enhancement. To accurately describe the
  ellipticity dependence of the long trajectory harmonics we must
  include a sub-cycle change of the initial velocity distribution of
  the electron and its excursion time. The resonant enhancement is
  observed only for the long trajectory contribution of a particular
  harmonic when a window resonance in argon, which is off-resonant in
  the field-free case, is shifted into resonance due to a large
  dynamic Stark shift.

\end{onecolabstract}
}]

The process of high-order harmonic generation
(HHG)\cite{McPherson1987JotOSoAB,FerrayJPB1988} driven by a strong
infrared (IR) laser field interacting with a rapidly ionizing medium
is the main light source for the field of attosecond
science\cite{PaulScience2001,Hentschel2001N,Agostini2004,Krausz2009}. The
HHG process can be used to produce attosecond pulses because there is a
natural, sub-cycle electron dynamics built into the physics of
HHG\cite{Haensch1990,Farkas1992}, which leads to a
very broad plateau of emitted harmonics. This means that studying the HHG
process itself in detail can, in principle, provide a deeper
understanding of strong field electron dynamics at the attosecond time
scale. Over the last decade
experiments\cite{BakerSci2006,SmirnovaNat2009} have indeed shown that
the sub-cycle dynamics of HHG are encoded in the harmonic spectrum,
though extracting them is complicated because of the highly non-linear nature
of the process.

Much of the promise in using HHG to better understand strong field
physics at the sub-cycle level can be attributed to the effectiveness
of the simple, semi-classical three-step model commonly used to
describe the generation
process\cite{SchaferPRL1993,CorkumPRL1993}. In this model, an
electron is first tunnel ionized and then accelerated by a strong
laser field. If the electron is driven back to the vicinity of the ion
by the oscillating strong field, the accumulated energy may be emitted
as a photon\cite{SchaferPRL1993,CorkumPRL1993,LewensteinPRA1994} when
the electron and ion recollide. The sequence of ionization and return
times leading to a specific harmonic frequency is loosely referred to
as a trajectory because much of the physics can be understood by
considering classical electron trajectories in a strong laser field,
ignoring atomic effects after the ionization step and before the
return. Depending on when during the laser cycle the ionization
occurs, the electron will have different excursion and return times to
the ion leading to different photon emission frequencies,
resulting in a comb of odd harmonics of the laser field if the process
is repeated over many laser cycles. 
Even in this simple model, however, there is not a one-to-one
correspondence between the harmonic emission strengths and specific
trajectories, because there are different trajectories leading to the
same final energy. Trajectories that lead to the same photon energy interfere at the single atom level.
The effect of this can either be studied\cite{ZairPRL2008,KolliopoulosPRA2014,PreclikovaArX2016} or circumvented, for example, through phase matching or spatial separation in the far field of the harmonics as is done in the present work.

Studying trajectory resolved contributions to the HHG spectrum is an
attractive proposition because different trajectories probe very
different ionization conditions and have different excursion
times. The most prominent contributions to the harmonic emission
strengths come from the so-called short and long trajectories, which
have excursion times of less than one laser cycle. Within a laser
cycle the long trajectories are ionized close to the peak field
strength and have an excursion time exceeding 0.65 laser cycles. The
short trajectories are ionized at low field strengths and have shorter
excursion times. Fortunately, the emission from these two trajectory
classes can be separated experimentally in the far field enabling, for
each harmonic frequency, comparison between ionization at two
different sub-cycle field strengths, followed by two different
excursion times. This requires, ideally, that high accuracy
measurements of both long and short trajectory contributions to each
harmonic be made in the same experimental setup\cite{CorsiPRL2006}.

Until recently most experimental efforts that make use of high
harmonics have been concentrated on optimizing HHG from short
trajectories, since their emission is well collimated and spectrally
narrow. In addition, trains of attosecond pulses have been
successfully created and measured by selecting the short trajectories'
contributions\cite{PaulScience2001,Lopez-MartensPRL2005}. The
emission from the long trajectories is more challenging to use because
it is spectrally broader and more divergent, hence it is usually
removed by spatial filtering and/or the selection of specific phase
matching conditions in experiments as they otherwise can affect the temporal structure of the attosecond pulses\cite{KrusePRA2010}. In this paper we report on
measurements made with very well-controlled, high repetition-rate
laser pulses, which allow us to make trajectory resolved HHG
measurements in argon gas while varying the ellipticity and the peak
field strength of the driving laser pulses. The results allow us to
elucidate new features in the sub-cycle ionization step that lead to
long trajectories, that is, ionization at high field strengths
followed by long excursion times.

In this article, we present two methods of probing sub-cycle
strong-field dynamics by comparing the trajectory resolved emission of
high harmonics and then studying the long trajectories in depth. In
the first part of the article, a detailed experimental comparison of
the ellipticity dependence as a function of harmonic order is
presented, for both the short and the long trajectories. While
harmonic generation using elliptically polarized driving fields has
been extensively studied for the short trajectories both
experimentally and
theoretically\cite{BudilPRA1993,BurnettPRA1995,SolaNP2006,ChangPRA2004,
  Strelkov2006PRA,Strelkov2012PRA,MoellerPRA2012}, the polarization dependence of the long
trajectories have so far only been investigated
theoretically\cite{Strelkov2006PRA,Strelkov2012PRA,ShilovskiPRA2008}. It
follows from the simple three-step model that harmonic generation will
be very sensitive to the ellipticity of the driving laser since the
field acting on the electron while it is far from the ion can cause it
to miss the recollision. Since the long and the short trajectories
have different excursion times, the impact of changing the ellipticity
will be different for the two classes of trajectories.

To explain the ellipticity dependence of the short trajectories it is
sufficient to include wave packet spreading due to quantum diffusion,
which we can model by including a distribution of momenta transverse
to the instanteneous field vector at the moment of ionization. This
distribution does not need to depend in detail on the moment of ionization,
since the ionization field strength is low for short trajectories. In
order to explain the ellipticity dependence of the long trajectories,
however, this simple quantum diffusion model is not enough. Due to a
larger variation in ionization field strength for the different long
trajectories, a field-strength dependent momentum distribution has to
be taken into account. We expect that at higher field strengths a
broader transverse momentum distribution results from the lowering of
the ionization barrier. We include this effect in our theoretical
analysis of the long trajectory data via a simple extension of the
three step model and find that it fits the our experimental data very well over the HHG
plateau.

In the second part of our study the sub-cycle sensitivity of
trajectory resolved HHG measurements is used to study a region of the
spectrum in which atomic resonances can alter the HHG signal. In
particular, a window resonance in argon that is far from any harmonic
of the laser frequency in the field-free case is shown to have a large
effect on the long trajectory harmonic closest to it, but little or no
effect on the short trajectory. We attribute this to the fact that the
long trajectory component is dynamically Stark shifted into resonance
by the laser field, which leads to a drastic enhancement of the
emission from the long trajectory, but not the short where the field
strength is much weaker and is not sufficient to shift the state into
resonance. We measure this effect for a set of resonant harmonics over
a range of driving field intensities.

\section*{Experimental setup}

The experimental setup used for the experiment presented in this
article is described in a recent publication\cite{LorekRSI2014} and
is briefly outlined here. An Yb:KGW based laser system (``Pharos'',
Light Conversion Ltd.) was used to deliver 170\,fs, pulses with a
central wavelength of 1030\,nm. The laser system has a variable
repetition rate between 1 and 600\,kHz, but all the presented data
were recorded at a repetition rate of 20\,kHz. The pulses were focused
tightly into a continuous argon gas jet, with a 90\,\micro{}m orifice,
using a 100\,mm focal length achromatic lens. Directly after the
interaction region, a differential pumping hole with an inner diameter
of 0.5\,mm was placed to minimize the background gas in the detection
chamber. The differential pump hole allowed for a pressure difference
of the background gas between the generation and detection chambers of
4--5 orders of magnitude. The HHG spectrum was measured using a
home-built imaging spectrometer based on a variable-line-spacing
grating and a microchannel-plate with an attached phosphor screen and
a camera with a resolution of 2456$\times$2058 pixels and a dynamic
range of 14 bits. The grating diffracts and refocuses the XUV in the
horizontal direction while the vertical direction is left
unaffected. Therefore the vertical direction provides the divergence
of the XUV light while the horizontal direction shows the spectrum.

\begin{figure}[tb]
  \begin{center}
    \includegraphics[width=0.9\columnwidth,trim = 1mm 155mm 10mm
    50mm,clip]{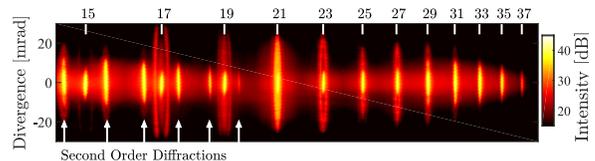} 
  \end{center}
  \caption{Typical harmonic spectra optimized to generate harmonics
    from both the short and the long trajectories.}
  \label{fig:experimentalsetup}
\end{figure}

\section*{Ellipticity measurement}

Figure~\ref{fig:experimentalsetup} shows a typical harmonic spectrum
when a linearly polarized driving laser is used and the gas jet is
placed in the focal plane of the generating beam. The experimental
parameters (pulse energy, gas density, spot size, etc.) were optimized
to generate harmonics from both the short and the long
trajectories.

The Gaussian transverse and temporal intensity profile of the driving laser, in
combination with the fact that the dipole phase of the long
trajectories has a stronger intensity dependence than the short
trajectories, result in a larger wavefront curvature and more
divergent light generated by the long
trajectories\cite{BelliniPRL1998,Lyngaa1999,GaardePRA1999}, this also explains the
spatial--spectral rings observed in the far field.  We therefore
attribute the inner part of the harmonic spectrum to be dominated by
the short trajectories, while the spatial--spectral rings are
attributed to interference between long trajectories of different
emitters. As the trajectory dependent dipole phase is strongest for
the low orders, the interference rings are mainly seen for the low end
of the plateau region. This spatial separation was exploited to study
the contributions from the long trajectories only.

\begin{figure}[tb]%
  \begin{center}
    \includegraphics[width=0.9\columnwidth,trim = 0 0mm 0 0mm,clip]{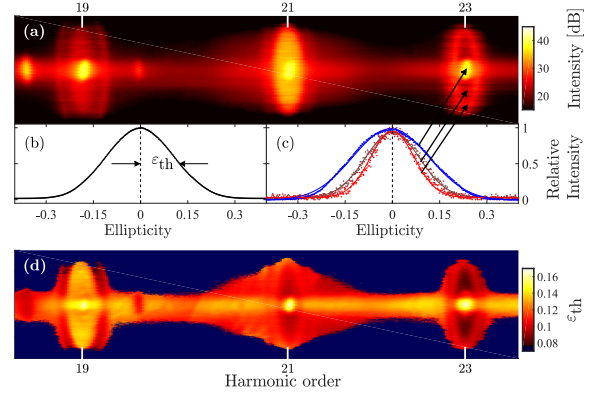} 
  \end{center}
  \caption{(a) Enlarged view of the harmonic spectrum at linear
    polarization for H19--H23 of Fig.~\ref{fig:experimentalsetup}. (b)
    Spatially and spectrally integrated signal of H23 as a function of
    ellipticity. (c) Measurement of the ellipticity dependence of
    three different spatial--spectral parts of H23 as indicated by the
    three arrows. The solid lines represent Gaussian fits to the
    experimental data.  (d) Pixel-by-pixel threshold ellipticity of
    the spatial--spectral region of part (a).}
  \label{fig:ellipticitydependence}
\end{figure}

A quarter-wave plate was used to introduce ellipticity, defined as the
ratio between the minor and major axis components of the driving laser
field. Figure~\ref{fig:ellipticitydependence}~(a) shows an enlarged
view of the spatial--spectral profile of harmonics 19--23 of
Fig.~\ref{fig:experimentalsetup}.
Figure~\ref{fig:ellipticitydependence}~(b) shows a measurement of the
spatially and spectrally integrated strength of harmonic 23 (H23) as a
function of ellipticity of the driving laser relative to the strength
at linear polarization.  The integrated signal clearly follows a
Gaussian distribution with respect to ellipticity as previously
observed\cite{BudilPRA1993}.

We define the threshold ellipticity, $\epsilon_\text{th}$, as the
amount of ellipticity required for the harmonic signal to drop by a
factor of two compared with linear polarization. Our very high
signal-to-noise ratio allows us to analyze the ellipticity dependence
of each pixel rather than the spatially and spectrally integrated
signal. Figure~\ref{fig:ellipticitydependence}~(c) presents the
strength of three different pixels within H23 as a function of
ellipticity. The strength of each pixel is fitted with a Gaussian
profile to extract the corresponding threshold ellipticity of each
pixel, which are used to create a two-dimensional map of the threshold
ellipticity as a function of energy and divergence angle. The full
threshold ellipticity maps for the conditions of
Fig.~\ref{fig:experimentalsetup} can be found in the Methods
section. Focusing on H23 in Fig.~\ref{fig:ellipticitydependence}~(d),
we observe three different regions of threshold ellipticity; an inner
region with a threshold ellipticity around 0.16, and two outer regions
with threshold ellipticities of around 0.09 and 0.1 respectively.

\begin{figure}[tb]
  \centering
  \includegraphics[width=\columnwidth,trim = 00mm 100mm 0mm 0mm,clip]{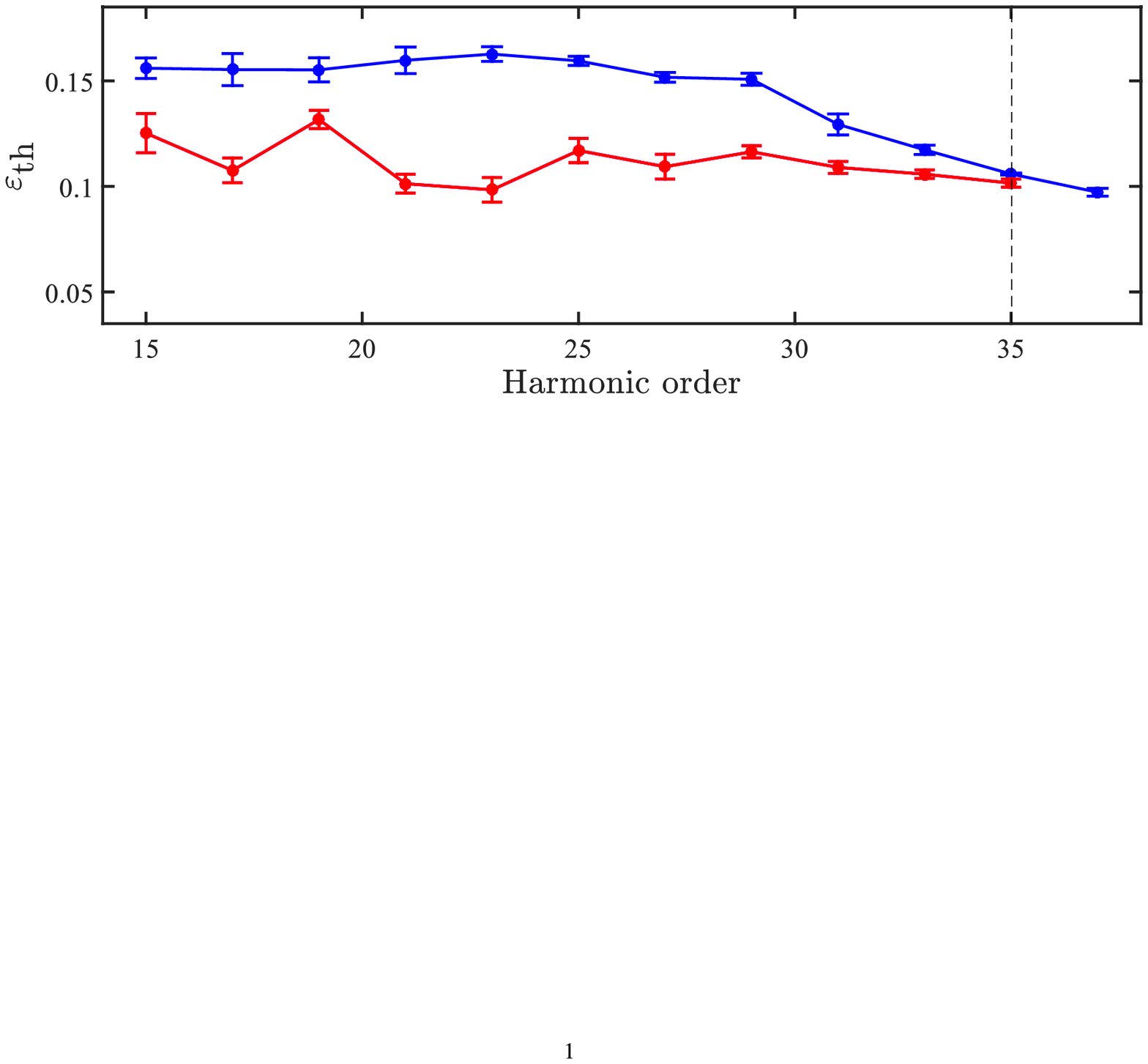} 

  \caption{The red (blue) line show the measured threshold ellipticity
    as a function of harmonic order for the long (short) trajectories
    when the laser is focused in the middle of the gas jet. The error
    bars indicate the standard deviations of the corresponding
    threshold ellipticity.}
  \label{fig:threshElliptTrajs}
\end{figure}
Figure~\ref{fig:threshElliptTrajs} presents the average threshold
ellipticity for both the long and the short trajectories as a function
of harmonic order for the conditions of
Fig.~\ref{fig:experimentalsetup}. For the short trajectories, we
observe a behavior similar to previous
measurements\cite{BudilPRA1993,SolaNP2006,MoellerPRA2012}, {\it
  i.e.}, the threshold ellipticity decreases slowly with increasing
harmonic order. A similar trend is also observed for the long
trajectories, in contrast to what would be expected if only the
excursion time of these trajectories is considered.

\section*{Ellipticity theory}

In a semi-classical model where the propagation step is calculated
classically and the electron only has a velocity parallel to the
electric field, only linearly polarized light would produce high-order
harmonics since any ellipticity will prevent the electron from
returning to its original position. The fact that high-order harmonics
are observed even for elliptically polarized light is usually
attributed to quantum diffusion; the electron wave packet spreads out
as it is accelerated in the laser field. The wave packet spread allows
for an overlap between the electron and the parent ion, even when the
electron is transversely displaced due to the elliptically polarized
laser field.

Quantum diffusion can be seen as resulting from an initial
distribution of velocities of the electron -- the more confined the
electron is in one direction, the more it will spread. In particular,
a spatial confinement in the direction perpendicular to the laser
field, will lead to a transverse velocity distribution which is
necessary for HHG. A rough estimate of the confinement is given by the
size of the groundstate. Using this estimate, a trajectory spending
longer time in the continuum will diffuse more which results in a
lower HHG yield.

For the short trajectories, the above estimate of the quantum
diffusion, which is independent of the ionization time, is sufficient
to explain the increase in sensitivity, as a function of harmonic
order. For this set of trajectories the highest energy photons are
produced by electrons with the longest excursion time. As the
transverse displacement of the electron at the point of recombination
increases with the excursion time, the trajectories leading to the
higher harmonics are displaced more than those leading to the low
orders. Therefore the overlap between the ion and the electron at the
recombination time decreases with harmonic order.

For the long trajectories this effect leads to the opposite result, as
the kinetic energy of the returning electrons decreases with
increasing excursion time.  To understand the experimentally observed
ellipticity dependence of this set of trajectories, we apply a model
that also takes the sub-cycle variation of the initial electron
velocity distributions into account, as well as the change in
excursion time for the different trajectories as the ellipticity is
varied. This effect plays a major role in the initial velocity
distribution as the long trajectory electrons are ionized closer to
the peak of laser field, where the atomic potential is more distorted
in the direction of the laser field, and thus the electron wave packet
is more perpendicularly confined at the time of
ionization\cite{TorlinaPRA2012}. The perpendicular confinement of the
electron at the ionization time leads to a large uncertainty in the
perpendicular velocity distribution.

Our method is similar to
\cite{Strelkov2006PRA,Strelkov2012PRA,MoellerPRA2012}, but the
definition of threshold ellipticity is not the same in the different
studies. The procedure is as follows:
First, we calculate the return energy of the electrons for the two
first sets of trajectories as a function of both ionization time and
ellipticity. The position of an electron released at time \(t_i\) in
an elliptical field
\(F(t)=F/\sqrt{1+\varepsilon^2}[\sin(\omega t);\varepsilon\cos(\omega
t)]\) is found by integrating the Newtonian equations of motion twice:

\begin{equation}
  \label{eq:electronPos}
  \begin{aligned}
    \vec{r}(t)=&\frac{F}{\omega²\sqrt{1+\varepsilon²}}
    \bmat{\sin(\omega
      t)-\sin(\omega t_i)-\omega(t-t_i)\cos(\omega t_i)\\\varepsilon[\cos(\omega t)-\cos(\omega
      t_i)+\omega(t-t_i)\sin(\omega
      t_i)]}\\
    +&
    (t-t_i)\vec{v}_i+
    \vec{r}_i,
  \end{aligned}
\end{equation}
where \(\vec{v}_i\) and \(\vec{r}_i\) are the initial velocity and
position, respectively; \(F\) is the field amplitude, \(\omega\) the
frequency of the fundamental field, \(\varepsilon\in[-1,+1]\) is the
ellipticity, with \(0\) meaning linear polarization along the \(x\)
axis. Atomic units are used.  We assume that
\(\vec{r}_i=\vec{r}(t_r)=0\), where $t_r$ is the moment of return.
Finding this time requires solving the transcendental equation
numerically. For elliptical polarization, the drift acquired by the
electron can be countered by an initial velocity \(\vec{v}_i\) that is
transverse to the driving field at the time of ionization (this is
analogous to quantum diffusion of the electron wavepacket as it is
accelerated in the laser field). Thus, we solve \eqref{eq:electronPos}
for \(t_r\) and \(\vec{v}_i\) for each \(t_i\in[0.25T,0.5T]\), \(T\)
being the period, and each \(\varepsilon\in[0,1]\).  We assume that
\(\vec{v}_i=\vec{v}_\parallel+\vec{v}_\perp\), where the two
components are parallel and perpendicular to the driving field at the
time of ionization. Furthermore, we assume that \(v_\parallel=0\),
such that all uncertainty is in the initial transverse momentum,
\(p_\perp= m_\mathrm{e}v_\perp\), (\(m_\mathrm{e}=1\) in atomic
units).

\begin{figure}[tb]
  \centering
  \includegraphics{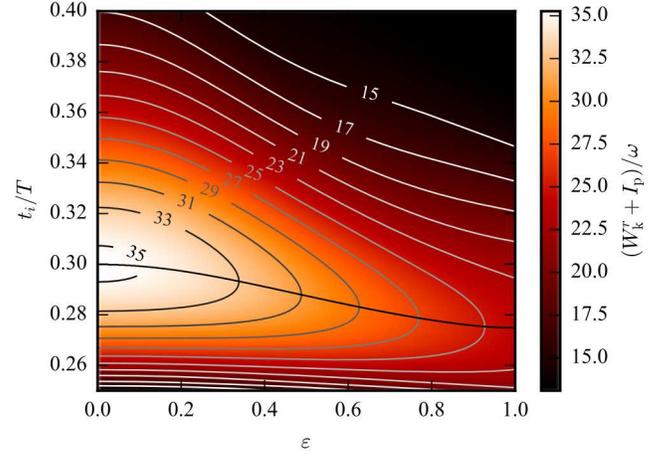}
  \caption{Map of return energies, as a function of ellipticity
    \(\varepsilon\) and ionization time \(t_i\). \(I_\mathrm{p}\) is
    the ionization potential of the ground state.  Plotted are also
    isoenergetic curves corresponding to the harmonics of the
    fundamental field, and the cut-off energy (shown by the solid black line), which decreases for
    increasing ellipticity. Trajectories, which are ionized earlier than the cut-off energy, correspond to the long trajectories. It is easy to see that a specific time of
    ionization does not correspond to a certain recombination energy.}
  \label{fig:returnEnergyMap}
\end{figure}

\begin{figure}[tb]
  \centering
  \includegraphics{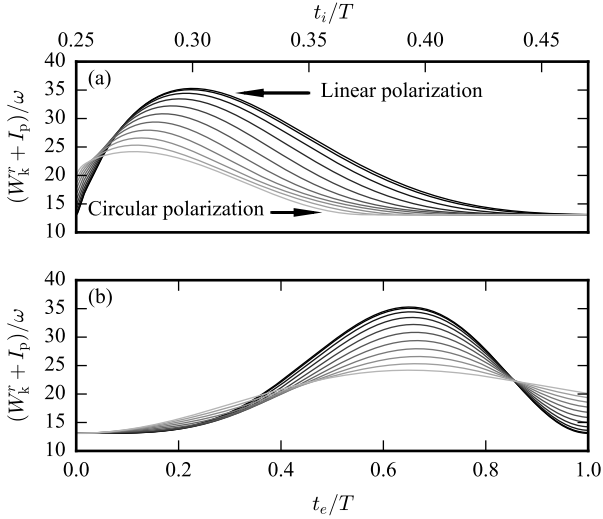}
  \caption{(a) Lineouts of Fig.~\ref{fig:returnEnergyMap} for 11
    equidistant ellipticities from linear to circular
    polarization. From this plot it is obvious that the cut-off
    decreases with increasing ellipticity and that it occurs for
    earlier ionization times. It is also clear that the initial
    timing necessary to produce a specific harmonic changes with the
    ellipticity. (b) The same as (a), but plotted as a function of
    excursion time \(t_e=t_r-t_i\) instead.}
  \label{fig:returnEnergyLineouts}
\end{figure}

The kinetic energy at the time of return is given by
$ W_{k}^r = p^2(t_r)/2$; this gives the map of energies seen in
Fig.~\ref{fig:returnEnergyMap}.
As can be seen in Fig.~\ref{fig:returnEnergyMap}, the cut-off position
is shifted when the ellipticity is increased (\emph{i.e.} the highest
energy photons can only be produced from linearly polarized light) and
the initial timing leading to a specific harmonic order is also
changed. This trend is even more clear when lineouts at different
ellipticities are presented as in Fig.~\ref{fig:returnEnergyLineouts}.

\begin{figure}[tb]
  \centering
  \includegraphics{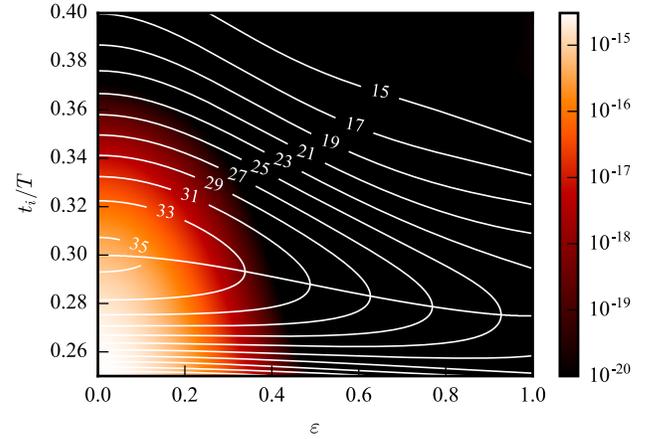}
  \caption{Map representing the combined probability of ionizing at
    time \(t_i\) and having the initial transverse velocity required
    for return as a function of ellipticity and ionization time. The
    isocurves are the same as in Fig.~\ref{fig:returnEnergyMap},
    representing constant return energy. Following an isocurve gives
    the probability of generating a certain harmonic, as a function of
    ellipticity.}
  \label{fig:transverseProbMap}
\end{figure}

The next step to estimate the harmonic yield is to calculate the
combined probability of ionizing at time \(t_i\)
and having the required initial velocity for
the electron to return. This is possible since
the correspondence between a certain harmonic and its ionization time
for different ellipticities is already calculated. Since the required
transverse momenta for the long trajectories to return are quite large
for high ellipticities, we use the full expression for the transverse
momentum distribution found in reference~\citenum{DeloneJOSA1991},

\begin{equation}
  \label{eq:transverseProb}
  w(p_\perp) \propto
  \exp\left[ -\frac{2(2I_{\textrm{p}}+p_\perp^2)^{3/2}}{3F(t_i)} \right].
\end{equation}
\(I_\mathrm{p}\) is the ionization potential of the ground state. The
tunneling rate is taken from ADK theory\cite{Ammosov1986SPJ}. This
combined probability, which is the product of the separate
probabilities described above, is displayed using a colour scale in
Fig.~\ref{fig:transverseProbMap}, as a function of ellipticity and
ionization time.
The isoenergetic curves from Fig.~\ref{fig:returnEnergyMap} are also
included in the figure. To calculate the yield of a given harmonic
order as a function of ellipticity, one extracts the probability along
the corresponding isoenergetic curve.

\begin{figure}[tb]
  \centering
  \includegraphics[width=0.9\columnwidth]{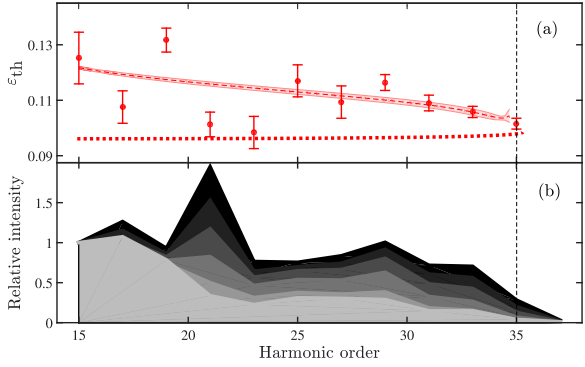}
  \caption{(a) The dashed lines show the numerically calculated
    threshold ellipticities for the long trajectories in the plateau
    region according to our model. Correspondingly, the dotted lines
    show the threshold ellipticity given by
    reference\cite{Strelkov2012PRA}. For comparison, the experimental
    data for the long trajectories are shown. The simulations were
    done for a laser intensity of $8.5\cdot 10^{13}$\,W/cm$^2$ and a
    wavelength of $1030$\,nm.  (b) Integrated harmonic spectra for
    in-focus generation with various ellipticities. The color scale
    corresponds to different ellipticities from black ($\epsilon = 0$)
    to light gray ($\epsilon = 0.25$) in steps of 0.05. The fillings
    between the different harmonics allow us to better visualize the
    differences. The various spectra have been normalized to the level
    of H15. The error bars used in (a) show the standard deviations of
    the threshold ellipticities of the corresponding harmonic.}%
  \label{fig:threshElliptInFocus}
\end{figure}

Finally, in the last step of our model, the calculated yield as a
function of ellipticity is fitted with Gaussian functions for each
harmonic to obtain the threshold ellipticities in a similar manner to
the experimental data. The result of the model for the long
trajectories is presented in Fig.~\ref{fig:threshElliptInFocus}~(a)
together with the experimental data.

Our extended model compares very well with the experiment presented in
this work; in particular the decrease in threshold ellipticity with
increasing harmonic order is explained. This is opposite to what would
be expected from sub-cycle field-independent quantum diffusion. It is
also opposite to the analytical expression presented in reference~\citenum{Strelkov2012PRA} which is included for comparison in
Fig.~\ref{fig:threshElliptInFocus}~(a) as a dotted line.

From Fig.~\ref{fig:threshElliptInFocus}~(a) it is clear that some of
the long trajectory harmonics (17, 21 and 23) have a lower threshold
ellipticity than what is predicted by the model. We attribute this to
the presence of atomic resonances in the vicinity of the corresponding
energies, which clearly cannot be captured by the model we are
using. In what follows, we demonstrate that these resonances can be
dynamically Stark shifted by the sub-cycle field strength and will
therefore influence the harmonic generation differently for the short
and the long trajectories. The spatial separation of the short and
long trajectories leading to the same energy, enable us to directly
compare the influence of the sub-cycle field strength. For a given
harmonic order, the short trajectory is initiated at a field strength
which is insufficient to shift the state into resonance, thereby
precluding the enhancement observed for the long trajectory initiated
at a higher sub-cycle field strength.

\section*{Resonant HHG}
\begin{figure}%
\begin{centering}
\includegraphics[width=1\columnwidth,trim=10mm 0mm 4mm 4mm,clip]{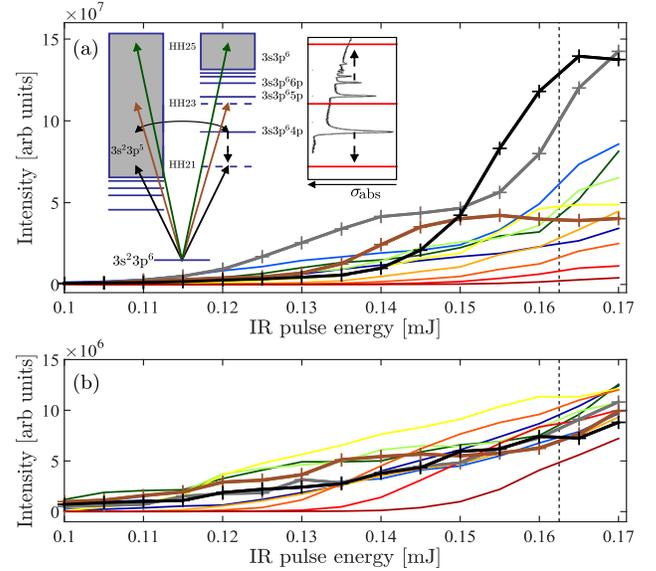}
	\end{centering}
  \caption{Experimentally measured long (a) and short (b) trajectory
    intensities of H15--H37 as a function of pulse energy of the
    fundamental field. H17 is shown in gray, H21 in black, and H23 in
    brown, while the remaining harmonics are shown in a rainbow color scale
    from dark blue (H15) to dark red (H37) with increasing order. The
    dashed line indicates roughly the conditions of
    Fig.~\ref{fig:experimentalsetup}. The inset of (a) schematically
    shows how the Rydberg series for the 3s electrons in combination
    with the continuum for the 3p electrons creates a series of window
    resonances. Absorption of an XUV photon can create a coherent
    superposition of the two valence electrons which interfere and
    affects the absorption cross-section
    $\sigma_\text{abs}$\cite{Fano1961PR}.  In argon this amounts to a
    reduction of the absorption
    cross-section\cite{SorensenPRA1994}. The vertical arrows within the inset indicate the expected direction of the light induced energy-shift of the respective 3s3p6np states.}%
  \label{fig:IntensityScan}%
\end{figure}
Resonant HHG in argon at a photon energy corresponding to H17 in our
experiment has previously been observed\cite{TomaJPB1999}. In our
study, we also see an effect for H21 and
H23~[Fig.~\ref{fig:threshElliptInFocus}~(b)]. We will focus the
discussion on H21, where the effect is most pronounced. At linear
polarization a strong enhancement of H21 is clearly observed, while
this enhancement is gone for an ellipticity of 0.2 as can be seen in
Fig.~\ref{fig:threshElliptInFocus}~(b). The change in ellipticity
leads to a variation in the intensity, since the pulse energy is kept
constant. This means that the observed effect can be due to either the
intensity or the ellipticity. In order to disentangle the two, an
intensity scan was performed for linearly polarized
light. Figure~\ref{fig:IntensityScan} shows the experimentally
measured intensities of the long (a) and short (b) trajectories for
H15--H37 as a function of IR pulse energy. As clearly observed, the
yield of H21 rises more rapidly for the long trajectories once the IR
pulse energy exceeds 0.14~mJ~[Fig.~\ref{fig:IntensityScan}~(a)], while
the emission from the short trajectories are left unaffected
[Fig.~\ref{fig:IntensityScan}~(b)].
 Enhancement of the long trajectories can also be clearly seen for H17 and H23, albeit, at slightly lower pulse energies.
Harmonic 17 qualitatively follows the trend predicted for the single atom response given in reference~\citenum{TomaJPB1999}, thus H17 is not further discussed in the present work. Harmonic 23 reaches a maximal strength at 0.15~mJ whereafter a slow decrease with respect to increased pulse energy is observed.

We interpret the behavior of H21 (but also H17 \& H23) to be the
result of HHG in the presence of an atomic resonance. Resonant HHG may
increase the harmonic yield through a number of different
mechanisms\cite{TomaJPB1999,TaiebPRA2003,GaneevOP2006,MilosevicJPB2007,
  StrelkovPRL2010,AckermannOE2012,GaneevPU2013}. For H21 with photon
energy of 25.2\,eV, the closest resonance is the
3s$^2$3p$^6 \rightarrow $ 3s$^1$3p$^6$4p$^1$ transition (26.6\,eV),
which is a window resonance\cite{Fano1961PR,SorensenPRA1994}.  Our
interpretation requires that the 3s$^1$3p$^6$4p$^1$ state, which is
lowest state in the 3s$\rightarrow$np closed channels, is red-shifted
by approximately 1.4\,eV [see inset in
Fig.~\ref{fig:IntensityScan}~(a)]. This is feasible as the dynamical
Stark shift of the 3s$^1$3p$^6$4p$^1$ state should be dominated by the
interaction with the 3s$\rightarrow$np closed channel, rather than
through coupling with the 3p$\rightarrow \Sigma$(s,d) open
channels\cite{WangPRL2010,OttNature2014}, and the Stark shifts on the
order of the ponderomotive energy are well
known\cite{BoerJPB1994,MevelJPB1992}. In addition to the Stark effect, the IR intensity also
causes a blueshift of the IR energy, and thereby the XUV photon energies. However, it was confirmed from the intensity scan that this effect is too small to explain the results, as the central frequencies of the harmonics did not change. Since we only
observe the enhancement for the long trajectory, we interpret this as
an effect of the comparably higher field strength for this trajectory,
at the time of ionization.

Reshaping of the argon HHG spectrum by this particular window
resonance has previously been observed\cite{RothhardtPRL2014},
however, the interpretation is fundamentally different in the work
presented here. In reference~\citenum{RothhardtPRL2014}, a few-cycle pulse was used
to generate broadband harmonics by HHG in a gas jet. The backing
pressure for the continuous gas jet was then increased significantly
so that all other wavelengths than exactly the resonant wavelength
were suppressed by re-absorption in the generating gas. This led to a
narrowing of H17 (of 800~nm) from a width of roughly 1.5\,eV to a
width comparable to the field-free width of the window
resonance\cite{SorensenPRA1994}. This leads us to believe that the
effect observed in reference~\citenum{RothhardtPRL2014} happens over a large
volume, where the IR intensity is weak. The results presented in this
article, however, is clearly an effect that take place at high laser
intensity, where the dynamical Stark effect is strong.

The behavior of H23 with respect pulse energy of the driving laser can be understood as an effect of over-shifting of the atomic resonances causing the enhancement. 
As indicated in the inset of Fig.~\ref{fig:IntensityScan}~(a) the field-free detuning of H23 is less than the detuning of H21 with respect to both the transition energies into the 3s$^1$3p$^6$4p$^1$ and the 3s$^1$3p$^6$5p$^1$ states, so 
the minimal required pulse energy for enhancement of this harmonic is lower. Nevertheless, the maximal enhancement factor is largest for H21, due to the strong dipole coupling with the red-shifted 3s$^1$3p$^6$4p$^1$ state. 

As the pulse energy is increased beyond the optimum energy for resonant generation of H23 the enhancement of this harmonic starts to vanish. The slow decrease likely originates from the long pulse duration of the driving laser, which means that a number of cycles will have the optimum energy shift. A similar effect is expected to occur for H21 at higher pulse energies, however, due to limitations of the laser system this was not seen in the present work.

Apart from the major effects on H17, H21 and H23 observed both in the ellipticity and the intensity measurements a minor amount of enhancement of H19, H25 and H27 can be observed for the long trajectories once the IR pulse energy exceeds 0.15\,mJ~[Fig.~8~(a)]. The field-free detuning from the 3s3p6np manifold of resonances are larger for these harmonics, so any enhancement effect on these harmonics is both expected to be less, and to occur at higher pulse energies in full agreement with the observation.

In conclusion, we have experimentally investigated the ellipticity and
intensity dependencies of HHG from the long and the short
trajectories. This type of measurements enables us to probe the
influence of the sub-cycle field strength on HHG process.  We have
shown that the well-established semi-classical model has to be
extended by taking the instantaneous field strength into account, to
also describe the general behavior of the long trajectories. We have
demonstrated how off-resonant states embedded in the continuum can
enhance long trajectory harmonics by being shifted into resonance by
the strong driving laser, different amounts for different trajectories
due to the sub-cycle nature of the generation process.  When the
driving laser field is strong enough to cause an enhancement at linear
polarization, these harmonics show a stronger ellipticity dependence
as the dynamical Stark shift depends on the polarization.

This study highlights the importance of systematical studies of the
generation process under various conditions. Furthermore, the
extension of the knowledge of the harmonic generation process to the
long trajectories will be beneficial for high-order harmonic
spectroscopy studies.

\begin{addendum}
\item
  The authors acknowledge fruitful discussions with P.~Johnsson,
  V.~V.~Strelkov, and M.~B.~Gaarde and thank\\ S.~L.~S\"{o}rensen for
  the cross-section data for argon. This research was supported by the
  Swedish Foundation for Strategic Research, the Marie Curie program
  ATTOFEL (ITN), the European Research Council (PALP), National Science Foundation Grant number: PHY-1307083,
  the Swedish Research Council and the Knut and Alice Wallenberg
  Foundation.
\item[Competing Interests] The authors declare that they have no
  competing financial interests.
\item[Correspondence] Correspondence and requests for materials should
  be addressed to E.W.L. (\texttt{esben.witting-larsen@fysik.lth.se}) or
  J.M. (\texttt{johan.mauritsson@fysik.lth.se}).
\item[Authors contributions]

  J.M. designed the experiment, E.L, C.M.H., E.W.L., A.L.H and J.M.
  designed and built the experimental apparatus. D.Z. and
  D.P. provided the laser system.  E.W.L., E.L. and S.C. conducted the
  experiments. E.W.L. performed the experimental data analysis and
  interpretation. S.C., E.W.L., K.J.S and J.M. developed the
  semi-classical modeling.  E.W.L., S.C, K.J.S., and J.M. wrote major
  parts of the manuscript.  All authors contributed to the discussion
  of the results and commented on the manuscript.

\end{addendum}
\begin{methods}
  \subsection*{Evaluation of experimental data}
  In this section we present the details of the analysis method used for
  the experimental data.

  It is well-known from the strong field approximation that for
  harmonics in the plateau region, there are several electronic
  trajectories, which may contribute to the generation process. Emission
  from these different trajectories interferes and shapes the far field
  spatial spectral profile. The phase of these trajectories can be
  approximated with a phase proportional to the intensity $I(x,y,z,t)$
  such that
  
  \begin{equation}
    \phi_q^{\textrm{traj}} = \alpha_q^{\textrm{traj}}I(x,y,z,t),
    \label{eq:}
  \end{equation}
  where $\phi_q^{\textrm{traj}}$
  is the trajectory dependent dipole phase and
  $\alpha_q^{\textrm{traj}}$
  is the proportionality constant. \ The first two sets of electron
  trajectories are usually referred to as the short and long
  trajectories. It well-established that in the plateau region the
  proportionality constants are much larger for the long trajectories
  than for the short trajectories\cite{GaardePRA1999,VarjuJMO2005}.

  The short trajectories can be isolated in the generation process by
  placing the gas jet behind the focal plane and adjusting gas
  pressure and pulse energy
  accordingly\cite{LHuillierPRL1993,SalieresSci2001}. A spectrum
  optimized for this is shown in
  Fig.~\ref{fig:fullthresholdellipmaps}~(a).
  Figure~\ref{fig:fullthresholdellipmaps}~(b) shows the corresponding
  threshold ellipticity map, which is extracted in a similar manner as
  in the main article.  When the gas jet instead is placed at the
  focus of the laser both sets of trajectories can efficiently be
  phase-matched by adjusting the other experimental parameters
  accordingly. Figure~\ref{fig:fullthresholdellipmaps}~(c) shows a
  spectrum optimized to generate with both sets of trajectories, while
  Fig.~\ref{fig:fullthresholdellipmaps}~(d) is the corresponding
  threshold ellipticity map.  As a consequence of the larger dipole
  phase of the long trajectories, the light produced by the these
  trajectories are more divergent. This effect was used to spatially
  separate the contributions from only the long trajectories in
  Fig~\ref{fig:fullthresholdellipmaps}~(c,d).
  \begin{figure}[tb]%
    \centering
		    \includegraphics[width=1.05\columnwidth,trim=0mm 88.7mm 5mm 00mm,clip]{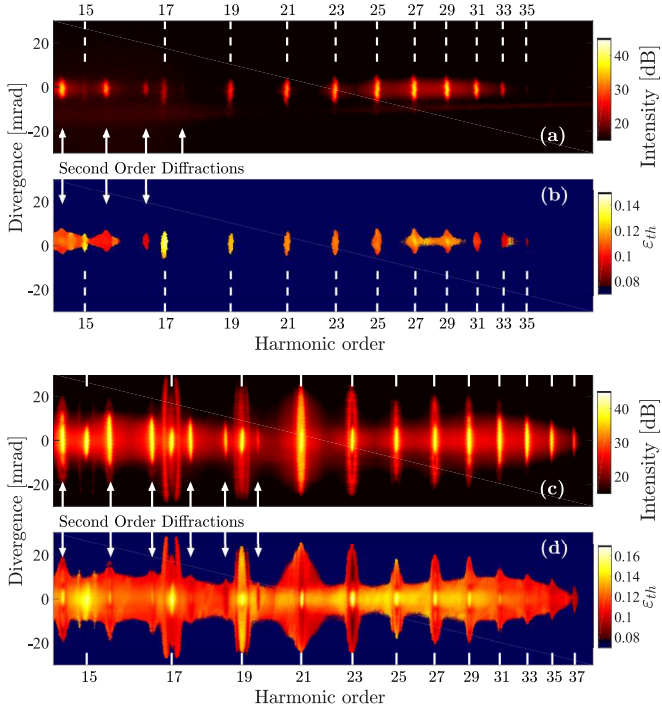} \\  

    \caption{(a) Harmonic spectrum at linear polarization when the gas
      jet is placed behind the focal plane of the laser. (b) Threshold
      ellipticity map as a function of energy and divergence angle for
      out-of-focus generation. (c) Harmonic spectrum at linear
      polarization when the gas jet is placed in the focal plane of the
      laser. (d) Threshold ellipticity map as a function of energy and
      divergence angle for in-focus generation. }
    \label{fig:fullthresholdellipmaps}%
  \end{figure}

  \begin{figure}[tb]%
	\begin{centering}
    \includegraphics[width=\columnwidth,trim=5mm 75mm 115mm 15mm,clip]{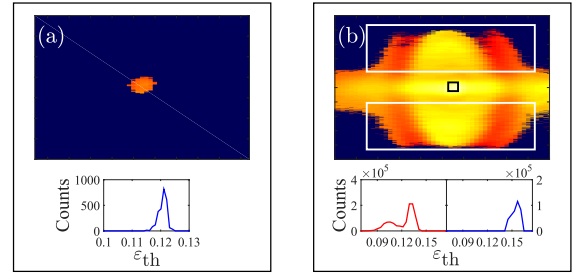}
			\end{centering}

    \caption{(a) An enlarged view of the ellipticity map of H19 for
      out-of-focus generation
      [Fig.~\ref{fig:fullthresholdellipmaps}~(b)]. The lower panel shows a
      histogram of the threshold ellipticities. Each pixel is weighted
      with the corresponding pixel strength at linear polarization.  (b)
      An enlarged view of the ellipticity map of H19 for in-focus
      generation [Fig.~\ref{fig:fullthresholdellipmaps}~(d)].  The lower
      panels show histograms of the threshold ellipticities within the
      white boxes (left panel) and the black box (right panel). The
      histograms are weighted with the corresponding pixel strength at
      linear polarization.}%
    \label{fig:histograms}%
  \end{figure}
  Figure~\ref{fig:histograms} shows enlarged views of H19 for
  out-of-focus generation [(a)] and in-focus generation [(b)]. In the
  out-of-focus case the harmonic exhibits a homogeneous
  spatial-spectral dependence with respect to ellipticity, this is not
  observed for the in-focus case, where several regions of ellipticity
  dependence are clearly observed.

  The homogeneity of the ellipticity dependence for out-of-focus
  generation [Fig.~\ref{fig:histograms}~(a)] reveal that in order to
  study ellipticity dependence of the short trajectories out-of-focus an
  imaging spectrometer is not needed and a spatial-spectral integration
  with respect to harmonic order would be sufficient. This is clearly
  not the case of in-focus generation [Fig.~\ref{fig:histograms}~(b)].

  Figure~\ref{fig:ThresholdellipLongShort} shows the threshold
  ellipticity as function of harmonic order for the conditions of
  Fig.~\ref{fig:fullthresholdellipmaps}.
  Figure~\ref{fig:ThresholdellipLongShort} shows normalized threshold
  ellipticity histograms as function of harmonic order. In the in-focus
  generation case both the on-axis emission and off-axis emission are
  shown, in (b) and (c) respectively, while for out-of-focus case only
  the on-axis emission is shown in (a). We note that for harmonics close
  to the cut-off indications of the long trajectory appears also for the
  out-of-focus case. In order to extract expectation values and standard
  deviations for the various threshold ellipticities as a function of
  harmonic order and trajectories the experimental data is smoothened
  using the Kernel density estimation
  method\cite{rosenblattAMS1956,parzenAMS1962}. After smoothing the
  data was fitted with two Gaussian distributions for the long
  trajectories, while the short trajectories where fitted with a single
  Gaussian distribution. The expectation value of the fitted Gaussian
  distributions are plotted as solid lines in
  Fig.~\ref{fig:ThresholdellipLongShort}, while the uncertainty bars
  show the corresponding standard deviations of the fits.

  \begin{figure}[tb]
    \includegraphics[width=\columnwidth,trim=25mm 10mm 35mm 10mm,clip]{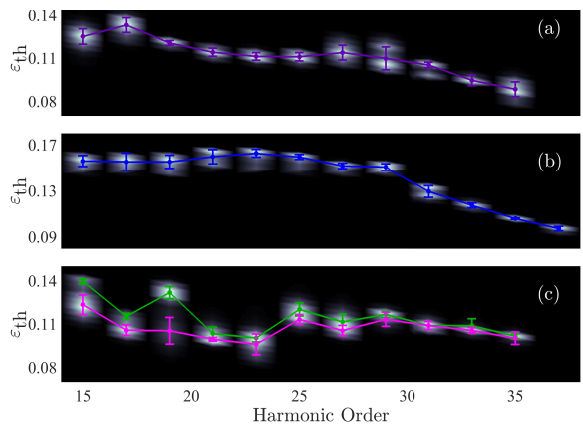}
    \caption{Normalized threshold ellipticity histograms as a function of
      harmonic order together with the extracted expectation values and
      standard deviations for: (a) The center of H15--H33 for out-of-focus
      generation. (b) The center of H15--H37 for the in-focus
      generation. (c) The outer regions of harmonic H15--H35 for in-focus
      generation.}
    \label{fig:ThresholdellipLongShort}%
  \end{figure}

  \subsection*{Detection efficiency}
  We measured the ratio in detection efficiency between horizontal and
  vertical polarization to be 1.38. 3D-TDSE calculations\cite{Strelkov2012PRA} show that the plateau harmonics exhibit a
  smaller ellipticity than the driving infrared laser. We therefore
  estimate that the upper limit of the non-fixed major axis impact on
  the measurement to be given by the following expression:
  
  \begin{equation}
    I_\text{det} \propto E_a^2 D_a+E_b^2 D_b,
    \label{eq:detection_eff}
  \end{equation}
  where $E_a$ and $E_b$ are the major and minor axis component of the
  infrared electrical fields, and $D_a$ and $D_b$ are the respective
  detection efficiencies. Using this expression together with the
  standard Jones matrix calculus for polarization of the infrared
  light we estimated the upper limit on the determination of the
  threshold ellipticity to be less than the presented standard
  deviations. The presented data was performed around the linear
  polarization direction with the highest detection
  efficiency. Therefore the threshold ellipticity might be
  systematically underestimated slightly.
\end{methods}

\end{document}